\documentclass{svjour3}                     
\smartqed  
\usepackage{graphicx}
\usepackage{mathptmx}      
\usepackage{natbib}
\usepackage{txfonts}
\usepackage{longtable}
\usepackage{lscape}
\usepackage{rotating}
\usepackage{multirow}
\usepackage{array}
\usepackage{wrapfig}

\def\deg{$^{\circ}$}
\journalname{SSRv}
\begin{document}

\title{The Solar Wind Charge-eXchange contribution to the Local Soft X-ray Background
}
\subtitle{Model to data comparison in the 0.1-1.0 keV band }


\author{Dimitra Koutroumpa \and
Rosine Lallement \and
Vasili Kharchenko \and
Alex Dalgarno}


\institute{D. Koutroumpa \at
              UMR 7620, IPSL/Service dA\'eronomie, CNRS, Universit\'e Pierre et Marie Curie, 
Universit\'e Versailles-Saint-Quentin, Verrières-le-Buisson, France \\
              Tel.: +33-1-64-47-42-94\\
              Fax: +33-1-69-20-29-99\\
              \email{dimitra.koutroumpa@aerov.jussieu.fr}           
          \and
           R. Lallement \at
              UMR 7620, IPSL/Service dA\'eronomie, Verrières-le-Buisson, France 
          \and
           V. Kharchenko \at
              Harvard-Smithsonian Center for Astrophysics, Cambridge, MA, USA
          \and
           A. Dalgarno \at
              Harvard-Smithsonian Center for Astrophysics, Cambridge, MA, USA
}

\date{Received: date / Accepted: date}

\maketitle

\begin{abstract}
The major sources of the Soft X-ray Background (SXRB), besides distinct structures as supernovae and superbubbles (e.g. Loop I), are: 
(i) an absorbed extragalactic emission following a power law, (ii) an absorbed thermal component ($\sim$2$\times$10$^6$ K) 
from the galactic disk and halo, (iii) an unabsorbed thermal component, supposedly at 10$^6$ K, attributed to the Local Bubble and 
(iv) the very recently identified unabsorbed Solar Wind Charge-eXchange (SWCX) emission from the heliosphere and the geocorona.

We study the SWCX heliospheric component and its contribution to observed data. In a first part, we apply a SWCX heliospheric simulation 
to model the oxygen lines (3/4 keV) 
local intensities during shadowing observations of the MBM 12 molecular cloud and a dense filament in the south galactic hemisphere with 
Chandra, XMM-Newton, and Suzaku telescopes. In a second part, we present a preliminary comparison of SWCX model results with ROSAT 
and Wisconsin surveys data in the 1/4 keV band. 

We conclude that, in the 3/4 keV band, the total local intensity is entirely heliospheric, 
while in the 1/4 keV band, the heliospheric component seems to contribute significantly to the local SXRB intensity and has potentially 
a strong influence on the interpretation of the ROSAT and Wisconsin surveys data in terms of Local Bubble hot gas temperature.

\end{abstract}

\section{Introduction}
\label{intro}
The surprising discovery of X-ray emission from comets \citep{lisse96sci} led to establishing Charge-eXchange (CX) collisions 
between highly charged Solar Wind ions and solar system neutrals as a very efficient non-thermal mechanism of soft X-ray emission \citep{cravens97}. 
\citet{cox98} suggested that X-ray emission induced in Solar Wind Charge-eXchange (SWCX) collisions with interstellar (IS) neutrals flowing across 
the Heliosphere would be an additional component of the diffuse soft X-ray background (SXRB).

Until that discovery, the SXRB was generally assumed to consist of an absorbed extragalactic emission following a power law, 
an absorbed thermal component ($\sim$2$\times$10$^6$ K) associated with the galactic disk and halo, and an unabsorbed thermal component, 
supposedly at 10$^6$ K, attributed to the Local Bubble \citep[][and references within]{mccammon90,kuntz00}. The Local Bubble (or Local Cavity) 
is defined as the region within $\sim$100 pc of the solar system particularly devoid of dense (and cold) gas as suggested by 
absorption lines in nearby star spectra \citep{lall03LB}. 


The 1/4 keV background generally presents a negative correlation with the column density of Galactic neutral hydrogen (dominated 
by a Galactic plane to-pole variation) that strongly suggests an ``absorption" model, in which the spatial structure of the 
SXRB is produced by absorption of a distant X-ray component \citep[e.g., ][]{bowyer68,marshall84}. However, the strong absorption 
predicted in such models was not confirmed by observation, neither was the large energy dependence of the absorption, 
so different solutions were needed.

At the time, the model most consistent with the data was the so-called displacement model \citep{sanders77,snowden90a}. In this model, 
the hot gas is in the foreground (in the Local Hot Bubble) with respect to the absorbing cool-gas regions, and higher X-ray intensities 
are produced in directions where the cavity has a greater extent (towards the LB chimneys at high galactic latitudes) and thus, 
a greater emission measure of the plasma in it.  

Since the discovery of SWCX emission, it became clear very quickly that it should contaminate at an unknown level all soft X-ray observations and 
it should depend strongly on SW flux and abundance variations as confirmed by the Long Term Enhancements detected during the 
ROSAT all-sky survey \citep{snowden95} and recent X-ray observations \citep{SCK04,smith05,henley07both}. 
The debate concerned the amount of contamination, and how the SWCX emission level would interfere with the Local Bubble thermal emission, 
since the two are most probably the major components of the unabsorbed fraction of the SXRB. Post-CX geocoronal emission can also 
contribute, but only during very intense SW enhancements and for short and easily identified intervals \citep[][]{cravens01}. 

The two components, although due to completely different mechanisms (collisional excitation for thermal plasma emission and electron 
capture for CX emission), produce spectra in the same energy range (E $\leq$ 1.5 keV) that are frequently confused. 
\citet{cravens00hel} estimated that SWCX emission could be as much as the LB emission. 
Line ratios, though, are very sensitive to gas temperature (and abundance) for thermal emission. For example, a 10$^6$ keV plasma might be 
responsible for all IS diffuse emission at 0.25 keV, while at 2-4 $\times$ 10$^6$ K it might significantly contribute to the 3/4 keV diffuse 
emission as well \citep[][and references therein]{raymond88,mccammon90}. CX line ratios mainly depend on heavy SW ion relative abundances 
and collision energy \citep{beiers01,kharchenko01,wargelin08}. Detailed spectral comparison of SWCX and thermal emission needs a spectral resolution 
that should allow to resolve characteristic features as the O\scriptsize{ VII}\normalsize\ triplet at 0.57 keV and in general individual lines 
from the continuum. Such a comparison is beyond the scope of this paper. 

In this paper we mainly focus on the intensity contribution of the SWCX emission to the observed (foreground) data in the 0.1 - 1.0 keV range. 
In section \ref{sec:model} we briefly present the basic simulation model that calculates the SWCX emission. 

In section \ref{sec:3quarter} we analyse the simulation results for the 3/4 keV range and in particular the oxygen 
lines (O\scriptsize{ VII}\normalsize\ at 0.57 keV and O\scriptsize{ VIII}\normalsize\ at 0.65 keV). We present typical stationary maps 
of oxygen lines in section \ref{sec:maps} and in section \ref{sec:dynamic} we present the dynamic model we developed to account 
for short-scale temporal variations of SWCX emission , with an 
application on the MBM 12 Suzaku observation of February, 2006. Finaly, we conclude section \ref{sec:3quarter} with a summary of 
shadowing observations on which we applied our SWCX dynamic model and discuss the results in the 3/4 keV range (sect. \ref{sec:shadows}).

I section \ref{sec:1quarter} we present a preliminary analysis of SWCX simulation results in the 1/4 keV range that 
we compare to observation data from the Wisconsin sounding rocket and ROSAT satellite surveys.

Finally, in section \ref{sec:discussion} we close our paper with a general discussion on our results.

\section{SWCX model description}
\label{sec:model}
In \citet{kout06} we have presented the basic stationary model calculating the SWCX emission in the inner heliosphere. We calculate 
self-consistently the neutral H and He density distributions in the inner heliosphere (up to $\sim$100 AU), in response to solar gravity, 
radiation pressure and anisotropic ionization processes for the two neutral species. Ionization is mainly due to charge-exchange 
with SW protons for H atoms and to solar EUV photons for He atoms. We also consider the impact of CX on the solar wind ions distributions. 
This interaction is described in the following reaction:
\begin{equation}
X^{\,Q+} + [H, He] \rightarrow X^{\,*(Q-1)+} + [H^{\,+}, He^+] 
\end{equation}
The collision rate per volume unit  R$_{X^{\,Q+}}$ (cm$^{-3}$ s$^{-1}$) of X$^{\,Q+}$ ions with the neutral heliospheric atoms is given 
by the equation:
\begin{equation}
R_{X^{\,Q+}} (r)\, =\, N_{X^{Q+}}(r)\, \upsilon _r\, (\sigma _{(H,X^{\,Q+})}\, n_H(r) +\, \sigma _{(He,X^{Q+})}\, n_{He}(r) )
= R_{(X^{\,Q+},H)} (r) + R_{(X^{\,Q+},He)} (r)
\end{equation}
where $\sigma_{(H,X^{\,Q+})}$ and $\sigma_{(He,X^{\,Q+})}$ are the hydrogen and helium CX cross-sections, n$_H$(r) and n$_{He}$(r) 
are the hydrogen and helium density distributions respectively, $\bar \upsilon_r = \bar V_{SW} - \bar \upsilon_n \approx \bar V_{SW}$ the 
relative velocity between SW ions and IS neutrals in the inner heliosphere, and $N_{X^{Q+}}(r)$ is the self-consistent solution 
to the differential equation:
\begin{eqnarray}
\frac{dN_{X^{\,Q+}}}{dx}\, & =\, & -\, N_{X^{\,Q+}}\, (\sigma _{(H,X^{\,Q+})}\, n_H(x) +\, \sigma _{(He,X^{\,Q+})}\, n_{He}(x))\\\nonumber
& & +\, N_{X^{\,(Q+1)+}}\, (\sigma _{(H,X^{\,(Q+1)+})}\, n_H(x) +\, \sigma _{(He,X^{\,(Q+1)+})}\, n_{He}(x))
\end{eqnarray}
expressing the evolution of the density distribution of ion X$^{Q+}$ along SW streamlines due to production (from CX reactions of 
ion X$^{(Q+1)+}$) and loss terms.

Cross-section uncertainties are mainly due to instrumental systematic errors and most important to collision energy dependance of cross-sections. 
Detailed uncertainties for individual ions are not given in literature, but average uncertainties of $\sim$30\% at most are reported 
\citep[][]{wargelin08}.

Then, we establish emissivity grids in units of (photons cm$^{-3}$ s$^{-1}$):
\begin{equation}
\varepsilon _{i} (r)\,  = R_{(X^{Q+},H)}(r)\,Y_{(E_i ,H)} + R_{(X^{Q+},He)}(r)\,Y_{(E_i ,He)} 
\end{equation}
where Y$_{(E_i ,M)}$ is the photon emission probability 
of spectral line E$_i$ following CX with 
the corresponding neutral species M (H or He individually).
For any line of sight (LOS) and observation date, this spectral line is given by:
\begin{equation} \label{groundI}
\displaystyle I_{E_i}\, (LU) = \frac{1}{4 \pi} \, \int_{0}^{\,\sim 100 AU} \varepsilon _{i} (s)\, ds 
\end{equation} 
which defines the average level emission of the spectral line for the particular date and LOS, as well as the solar cycle phase 
(minimum or maximum) corresponding at this date. 

\section{SWCX in the 3/4 keV Band}
\label{sec:3quarter}
\begin{figure*}
\begin{center}
  \includegraphics[width=0.85\textwidth]{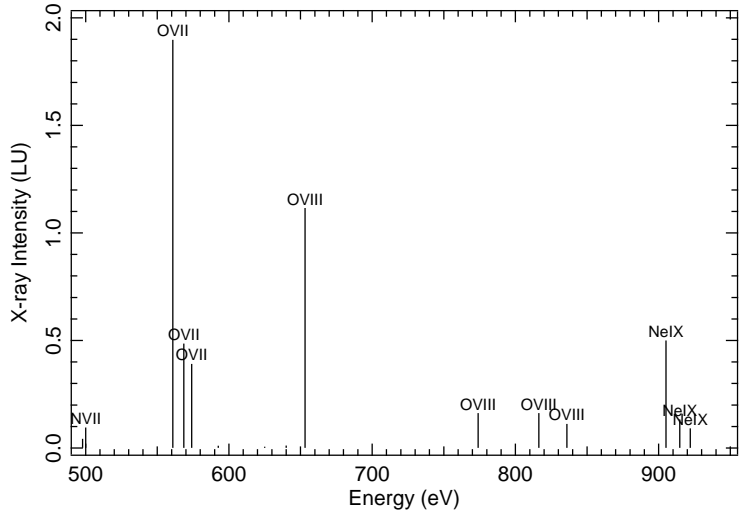}
\caption{SWCX model spectra in Line Units (photons cm$^{-2}$ s$^{-1}$ sr$^{-1}$) for the 3/4 keV band. The emitting ion is noted above each line.}
\label{fig:spec3_4}       
\end{center}
\end{figure*}

An example of calculated SWCX emission spectra in the 3/4 keV (0.5-1.0 keV) range is presented in figure \ref{fig:spec3_4}. The major 
emission lines contributing in this energy range are: \\
(i) the He-like O\scriptsize{ VII}\normalsize\ multiplet, consisting of the following transitions:\\
- the 2$^3$S$_1$ triplet at 560.9 eV, called forbidden (O6f)\\
- the 2$^3$P$_1$ triplet at 568.5 eV, called intercombination line (O6i)\\
- the singlet state 2$^1$P$_1$ at 574 eV, usually refered to as resonance line (O6r), and\\
(ii) the H-like O\scriptsize{ VIII}\normalsize\ Lyman-$\alpha$ line at 653.1 eV. \\
In this list we can also add the He-like Ne\scriptsize{ IX}\normalsize\ multiplet (905.1 eV, 914.7 eV and 922.1 eV), but actual 
observing instruments do not always allow the detection of this line.

\subsection{Stationary Model Maps}
\label{sec:maps}
\begin{wrapfigure}{r}{5.6cm}
    \begin{minipage}[l]{0.48\textwidth}
\includegraphics[width=\textwidth]{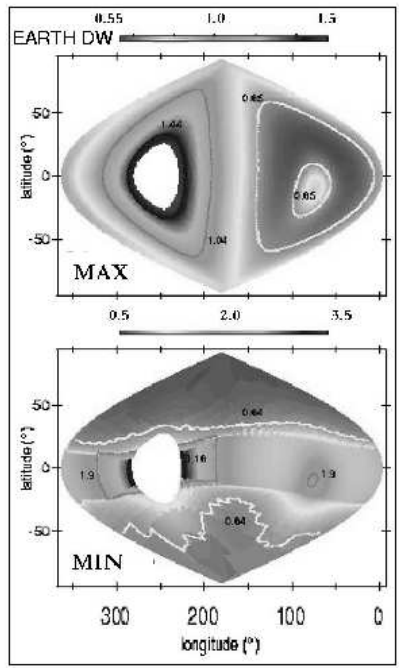}
    \end{minipage}\\
    \begin{minipage}[l]{0.45\textwidth}
\caption{Calculated full sky maps in ecliptic coordinates of the 0.65 keV O\scriptsize{ VIII}\small\ line for solar maximum (upper panel) 
and solar minimum (lower panel) conditions. The observer is situated downwind (ecliptic longitude 75$^\circ$). 
Colorscales are in units of 10$^{-9}$ erg cm$^{-2}$ s$^{-1}$ sr$^{-1}$. }
\label{fig:O7DW}       
   \end{minipage}\hfill
\end{wrapfigure}

In figure \ref{fig:O7DW} we present an example of calculated full-sky maps of heliospheric emission for line O\scriptsize{ VIII}\normalsize\ (0.65 keV) 
in ecliptic coordinates, on December, 5 (when the Earth is located downwind) of a typical solar maximum (upper panel) and solar minimum 
(lower panel) year. Colorscales represent O\scriptsize{ VIII}\normalsize\ line intensity given in units of 
10$^{-9}$ erg cm$^{-2}$ sr$^{-1}$ s$^{-1}$, the red colour corresponding to minimum and the blue to maximum values. 
We have removed from every map 
a data portion of 20\deg $\times$ 20\deg around the solar disk where no instrument can observe.

What we need to point out here are the striking differences between maximum and minimum solar conditions. At low latitudes, 
near the solar equator where slow SW dominates, X-ray emission is more intense for solar minimum than for solar maximum because 
the neutral atom content is higher during solar minimum. During solar minimum, photo-ionization is less efficient and H and He are less 
readily destroyed by photo-ionization. Since gravitational pressure exceeds radiation pressure, neutral H atoms have incoming 
trajectories that fill the ionization cavity.

On the other hand, at high ecliptic latitudes, LOS's are mostly affected by fast wind during solar minimum, and the X-ray emission 
is dominated by differences of ion relative abundances with respect to maximum cycle phase. Indeed, O$^{8+}$ is completely absent 
from the fast SW and the O\scriptsize{ VIII}\normalsize\ line is expected to be much fainter at solar minimum than solar maximum, 
as it is seen in the maps in fig. \ref{fig:O7DW}.

O$^{7+}$ has the same trend in relative abundance variations as O$^{8+}$ between solar minimum and solar maximum. Nevertheless, 
instead of being completely absent in the fast wind, it is only strongly depleted ([O$^{7+}$ / O] = 0.03) with respect to slow wind 
([O$^{7+}$ / O] = 0.2). Therefore, we expect that O\scriptsize{ VII}\normalsize\ line (0.57 keV) variations from solar maximum to solar minimum 
will also have the same trend as O\scriptsize{ VIII}\normalsize\ shown in the maps.

Two excellent examples of such differences between solar minimum and maximum conditions for the same LOS, are the MBM 12 and South Galactic 
Filament (SGF) shadowing observations, analysed by \citet{smith05,smith07} and by \citet{henley07xmm,henley07suz,henley07both} respectively. 
These fields were thoroughly analysed for their SWCX contamination in \citet{kout07} and the main outline and conclusions are presented 
in section \ref{sec:shadows}.

The MBM 12 LOS points toward galactic coordinates (159.2\deg , -34.47\deg ), which translates to (47\deg , 3\deg ) in helioecliptic 
coordinates, very close to the ecliptic plane and inside the limits of the equatorial slow wind zone. Therefore, during solar minimum 
the MBM 12 exposures (Suzaku) will yield higher SWCX contamination than solar maximum exposures (Chandra) in the 3/4 keV range.

In the case of SGF, at very high southern ecliptic latitude (353\deg , -73\deg ) the effect is the opposite compared to MBM 12, 
because it is dominated by differences in the solar ions relative abundances. Therefore, the SWCX oxygen emission in the 3/4 keV range for the SGF 
field is expected to be much fainter at solar minimum (Suzaku observations) than solar maximum (XMM-Newton observations).

\subsection{Dynamic Simulations}
\label{sec:dynamic}
In \citet{kout07} we presented a dynamic variant of the basic stationary SWCX model, used to simulate the heliospheric X-ray modulation 
due to SW proton flux and abundance variations. For each field simulated we use real-time flux and abundance measurements 
from {\em in situ} solar wind instruments (Wind, ACE-SWEPAM) and we model SW enhancements as simplified Corotating Interaction Regions (CIR) of a 
step function form. The total width of the CIR is defined such that the total duration of the step function is the same as the measured 
enhancement in SW instruments. The step function amplitude is defined so that its total area is equal to the integral of 
the measured flux during the same period of time.

At each instant we define the form of the CIR (as a Parker spiral) and its propagation in the interplanetary space, taking into account 
solid solar rotation (27-day period), the radial propagation speed, the ``ignition" time on the solar disk towards 
each radial direction, and the total width of the CIR. Only in the cases of CMEs, we neglect the solar rotation, 
since radial propagation is dominating the CME structure. Depending on the CIR's width and propagation instant, we calculate the fragment 
of the LOS affected by the CIR and the local emissivity $\varepsilon _i'$(r) modified by the SW proton flux and heavy ion abundance variations. 
We can then reproduce the temporal variation of the X-ray intensity levels during the periods of observation in simulated lightcurves 
for each of our targets. Abundance variations, can be correlated or anticorrelated with proton fluxes, so they can either emphasize 
or compensate for the influence of the SW proton flux enhancements.

\subsubsection{The Case of the MBM 12 Suzaku Observation of February 2006}
\label{sec:mbm12}
The observation of MBM 12 with Suzaku on February 3-8, 2006 was performed in two consecutive exposures, ON-CLOUD (3-6/02/2006, 
for a total of 231 ks) and immediately after that, OFF-CLOUD (6-8/02/2006, for a total of 168 ks) \citep{smith07}. A detailed 
description of the simulation we performed on this field is given in \citet{kout07}. 

As we already explained, the MBM 12 LOS is located at (47\deg , 3\deg ) of helioecliptic coordinates. In the February Suzaku observation 
this geometry is pointing directly toward the He cone, which is only $\sim$2 AU away from the observer, where the emissivity on the LOS 
is maximum. The observation geometry was presented in figure 1 of \citet{kout07}. The cone, with a denser distribution of helium atoms, 
is acting as an amplifier to any SW flux variation. 

\begin{figure*}
    \begin{minipage}[c]{0.68\linewidth}
  \includegraphics[width=\textwidth]{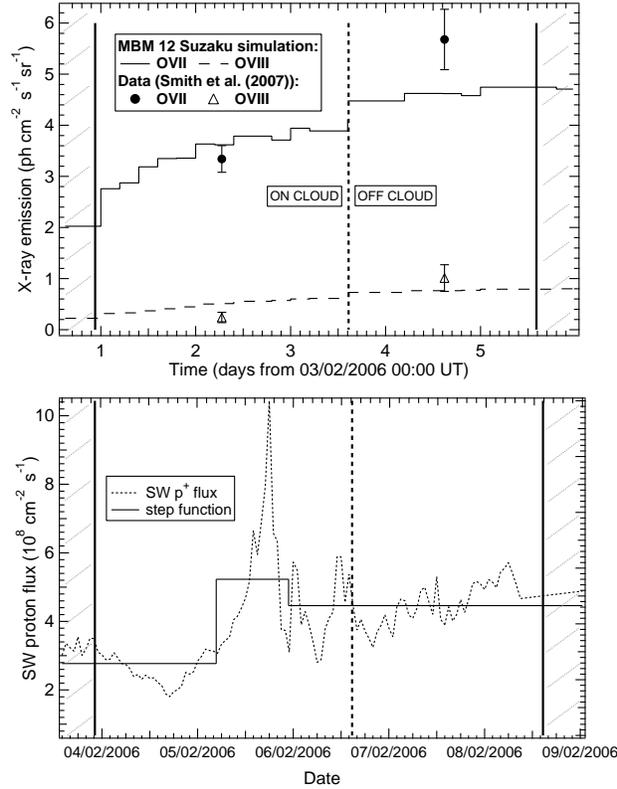}
    \end{minipage}\hfill
\begin{minipage}[c]{0.32\linewidth}
\caption{\textit{Top}: O\scriptsize{ VII}\small\ and O\scriptsize{ VIII}\small\ line simulated lightcurves (in LU) for the MBM12 ON 
and OFF observations with Suzaku during the period 03-08/02/2006. 
Plain line stands for the O\scriptsize{ VII}\small\ emission and dashed line stands for the O\scriptsize{ VIII}\small\ emission. 
Dots represent the measured ON and OFF O\scriptsize{ VII}\small\ line flux, and triangles represent 
the measured ON and OFF O\scriptsize{ VIII}\small\ line flux \citep[][]{smith07}.\newline \textit{Bottom}: 
Solar wind proton flux (dotted line) in units of $10^8$ cm$^{-2}$ s$^{-1}$ for the same period. The step function simulating 
the SW enhancement is presented by the plain black line. In both panels, the vertical plain lines represent the start 
and end of the observation period, while the dashed vertical line is the separation between the ON and OFF exposures.}
\label{fig:1}       
\end{minipage}
\end{figure*}

The MBM 12 Suzaku observation was influenced by a short SW perturbation recorded at the end of the ON-CLOUD pointing. 
We model this enhancement as a step function in particle flux of 5.2 $\times$ 10$^8$ cm$^{-2}$ s$^{-1}$ propagating at a low speed 
(V$_{SW}$ = 350 km s$^{-1}$) for 0.75 d. The SW proton flux remained at a high, but stable, level equal 
to 4.42 $\times$ 10$^8$ cm$^{-2}$ s$^{-1}$ after the spike. In the lower panel of figure \ref{fig:1} we present the measured SW flux curve 
with the dotted line and the modeled step function with the plain line.

The resulting simulated lightcurves for O\scriptsize{ VII}\normalsize\ and O\scriptsize{ VIII}\normalsize\ are shown in the upper panel 
of figure \ref{fig:1} in plain and dashed lines respectively. The plain vertical lines mark the limits of the total observing period 
and the dashed vertical line shows the limit between the ON and OFF cloud exposures. The dynamic model, taking into account SW measurements, 
is predicting a smooth but gradual rise of the oxygen line intensities in the ON and OFF exposures. The O\scriptsize{ VII}\normalsize\ 
and O\scriptsize{ VIII}\normalsize\ line fluxes averaged for the ON-CLOUD exposure yield 3.56 LU and 0.5 LU respectively, 
while averages on the OFF-CLOUD exposure yield 4.62 LU and 0.77 LU for O\scriptsize{ VII}\normalsize\ and 
O\scriptsize{ VIII}\normalsize\ respectively. The equivalent averaged Suzaku data for O\scriptsize{ VII}\normalsize\ and 
O\scriptsize{ VIII}\normalsize\ are shown in the figure with plain dots and hollow triangles respectively.

The model-data comparison for the ON-cloud exposure shows that LB emission is negligible, since the SWCX model accounts for all of the 
observed intensity. The X-ray flux increase ($\sim$30\% and $\sim$55\% for O\scriptsize{ VII}\normalsize\ 
and O\scriptsize{ VIII}\normalsize\ respectively) in the OFF-CLOUD exposure would change the data interpretation in terms 
of hot halo gas temperature, since \citet{smith07} assumed a constant foreground emission.

\subsection{Summary of Shadowing Observations}
\label{sec:shadows}
%
\setlength{\extrarowheight}{1.5mm}
\begin{table*}
\begin{center}
\caption{Comparison of shadowing observation foreground data to SWCX model for\, O\scriptsize{ VII} \small and O\scriptsize{ VIII} 
\small line intensities. $^a$ Observation date and observing instrument: C for Chandra, X for XMM-Newton and S for Suzaku. 
$^b$ MBM 12 Suzaku-off exposure is included in the analysis, as foreground emission was not constant (see \S \ref{sec:mbm12} for details). 
$^c$ Revised values for SGF foreground line intensities, from \citet{henley07both}.}
\label{tab:1}       
\begin{tabular}{llcccccc}
\hline\hline\noalign{\smallskip}
& & \multicolumn{3}{c}{O\scriptsize{ VII} \small (LU)} & \multicolumn{3}{c}{O\scriptsize{ VIII} \small (LU)}\\
\noalign{\smallskip}\hline\noalign{\smallskip}
Target & Observation $^a$& Data & Data $^c$ & SWCX & Data & Data $^c$ & SWCX \\
\noalign{\smallskip}\hline\noalign{\smallskip}
MBM 12 & Aug. '00, C & 1.79\,$\pm$\,0.55 & & 1.49 & 2.34\,$\pm$\,0.36 & & 2.13\\
MBM 12 & Feb. '06, S-on & 3.34\,$\pm$\,0.26 & & 3.56 & 0.24\,$\pm$\,0.10 & & 0.50 \\
MBM 12 & Feb. '06, S-off $^b$ & 5.68\,$\pm$\,0.59 & &4.62 & 1.01\,$\pm$\,0.26 & & 0.77\\
SGF & May, '02, X & 3.40 & 6.2$^{+2.8}_{-2.9}$& 3.16 & 1.00 & ... & 1.02 \\
SGF & Mar. '06, S & 0.13 & 1.1$^{+1.1}_{-1.4}$& 0.34 & N.C. & 1.0$\pm$1.1& 0.02\\
\noalign{\smallskip}\hline
\end{tabular}
\end{center}
\end{table*}

In Table \ref{tab:1} we resume the MBM 12 and South Galactic Filament shadowing observations analysed in detail in \citet{kout07}. All four observations were 
simulated with the SWCX model with SW conditions as close as possible to measurements in solar instruments, to account for the 
SWCX contamination of the oxygen line emission in each field. 

Besides the basic differences due to solar maximum or solar minimum conditions during the observing period for each field, that we reported 
in section \ref{sec:maps}, three of the four observations (MBM 12 Chandra, Suzaku and SGF XMM observations 
respectively) were found to be highly contaminated by short-term SWCX emission variations. In particular, in our simulations we confirmed 
the contamination of the MBM 12 Chandra observation by a strong CME as suggested by \citet{smith05}. We suggested that the SGF XMM observation 
was also highly contaminated, most probably by a CME, which was confirmed by \citet{henley07both} when they re-analysed and compared 
their Suzaku and XMM data. 

The observation data represent unabsorbed (attributed to the LB) O\scriptsize{ VII}\normalsize\ and O\scriptsize{ VIII}\normalsize\ 
line intensities as derived from the authors \citep{smith05,smith07,henley07xmm,henley07suz,henley07both}. 
In particular, for the SGF field we give two sets of data: from the initial analysis published by \citet{henley07xmm} and \citet{henley07suz} 
\citep[values we used in][]{kout07} and the revised values published by \citet{henley07both}.

The shadows block more or less efficiently the Galactic Halo oxygen emission, and in their analysis the authors derive 
the unabsorbed foreground emission of the oxygen lines, that we use to compare to the SWCX heliospheric emission. 
The only exception is in the Suzaku/MBM 12 observation \citep{smith07} where the O\scriptsize{ VII}\normalsize\ and 
O\scriptsize{ VIII}\normalsize\ halo emission is added to the foreground emission 
for the OFF-CLOUD exposure. Contrary to the constant foreground assumption in the \citet{smith07} analysis, we demonstrated 
that there is a $\sim$30\% and $\sim$55\% increase in the OFF-CLOUD simulated SWCX O\scriptsize{ VII}\normalsize\ 
and O\scriptsize{ VII}\normalsize\ line intensities respectively, due to the brief SW enhancement at the end of the ON-CLOUD exposure 
(see \S \ref{sec:mbm12}). This increase was erroneously attributed to the Galactic Halo emission, which should be revised, and thus we include 
these values in the analysis as well.

\begin{figure*}
    \begin{minipage}[c]{0.51\linewidth}
  \includegraphics[width=\textwidth]{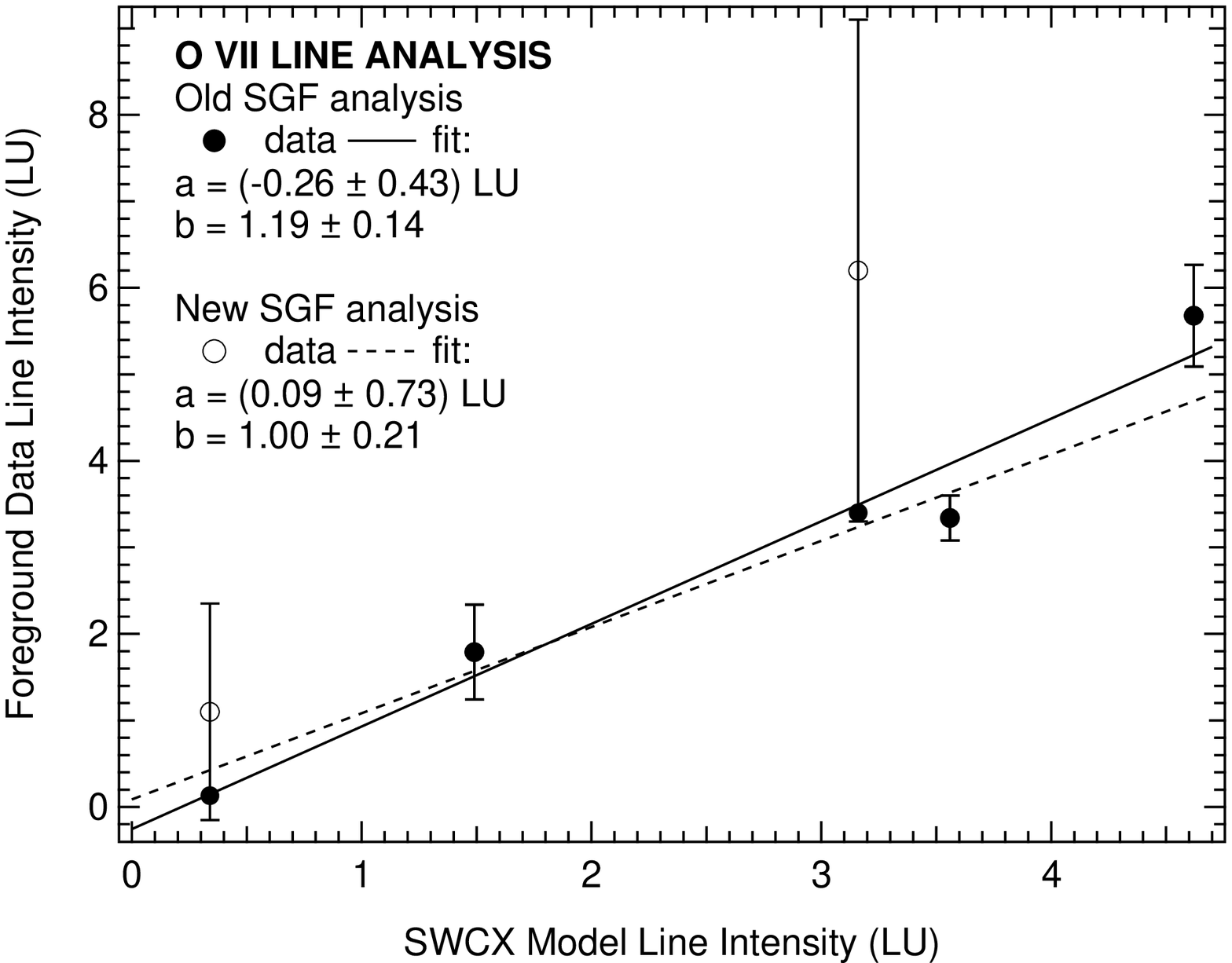}
    \end{minipage}\hfill
\begin{minipage}[c]{0.51\linewidth}
  \includegraphics[width=\textwidth]{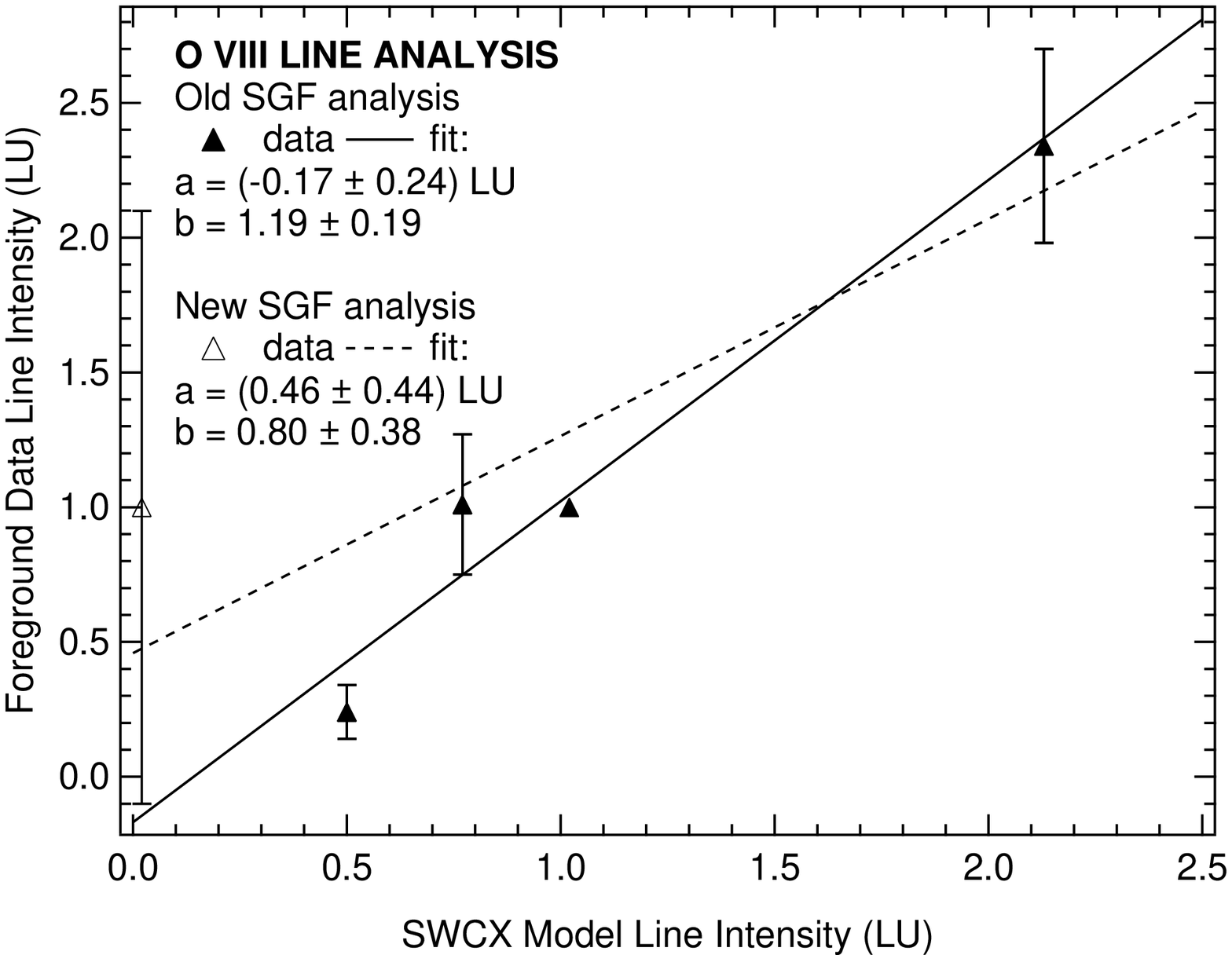}
\end{minipage}
\caption{Linear fits of oxygen line intensities. Foreground emission data from shadowing observations in litterature 
\citep[MBM 12:][]{smith05,smith07} and \citep[SGF:][]{henley07xmm,henley07suz} versus SWCX model results 
for the O\scriptsize{ VII}\small\ (left panel) and O\scriptsize{ VIII}\small\ (right panel) line intensities. All linear fits 
are calculated with no error-bar weighting except for the O\scriptsize{ VII}\small\ New SGF analysis fit (left panel, dashed line). 
See \S \ref{sec:shadows} for more details.}
\label{fig:newSHfit}       
\end{figure*}

In figure \ref{fig:newSHfit} we linearly fit the data over the SWCX model results for the O\scriptsize{ VII}\normalsize\ (left panel) 
and O\scriptsize{ VIII}\normalsize\ (right panel) lines for the initial (plain lines) and revised (dashed lines) data sets. 
We did not use data errorbars as standard deviation to weight the fit, because not all data errorbars were communicated 
\citep{henley07xmm,henley07suz}. The only exception is the O\scriptsize{ VII}\normalsize\ line fit, since \citet{henley07both} 
provided errorbars for both XMM and Suzaku O\scriptsize{ VII}\normalsize\ detections. The old and new fit coefficients (y(LU) = (a + bx)(LU)) 
for the O\scriptsize{ VII}\normalsize\ and O\scriptsize{ VIII}\normalsize\ lines are also noted on the figure.

The first linear fits (in plain lines in fig.\ref{fig:newSHfit}) we performed \citep{kout07} had a slope of 1.19 ($\pm$0.14; 0.19 at 1$\sigma$ 
for O\scriptsize{ VII}\normalsize\ and O\scriptsize{ VIII}\normalsize\ resp.) meaning 
that the SWCX model reproduced very well the local foreground emission measured in shadows. Moreover, the fit also gave an estimate 
of the residual foreground emission to be attributed to the LB. The foreground was found to be (-0.26 $\pm$ 0.43 at 1$\sigma$) LU for 
O\scriptsize{ VII}\normalsize\ and (-0.17 $\pm$ 0.24 at 1$\sigma$) LU for O\scriptsize{ VIII}\normalsize\, which means 
that with a high probability the LB O\scriptsize{ VII}\normalsize\ and O\scriptsize{ VIII}\normalsize\ emission 
is negligible compared to the heliospheric emission.

The new fits (in dashed lines, fig.\ref{fig:newSHfit}), where we substituted the SGF observed foreground data with the revised values 
of \citet{henley07both} give slightly different results, although the correlation to the simulated values remains very satisfying (the correlation 
is even improved for the O\scriptsize{ VII}\normalsize\ line intensities) and the residual LB emission is still consistent to zero.

\section{SWCX in the 1/4 keV Band}
\label{sec:1quarter}
In this section we present a preliminary study of the SWCX background in the 1/4 keV (0.1-0.3 keV) band that we compare with the ROSAT and 
Wisconsin surveys data.  In section \ref{sec:HM1/4} we describe briefly the SWCX simulation and in section \ref{sec:data} we compare the model results 
with the survey data sets.

\subsection{Simulations}
\label{sec:HM1/4}

\begin{figure*}
  \includegraphics[width=1.\textwidth]{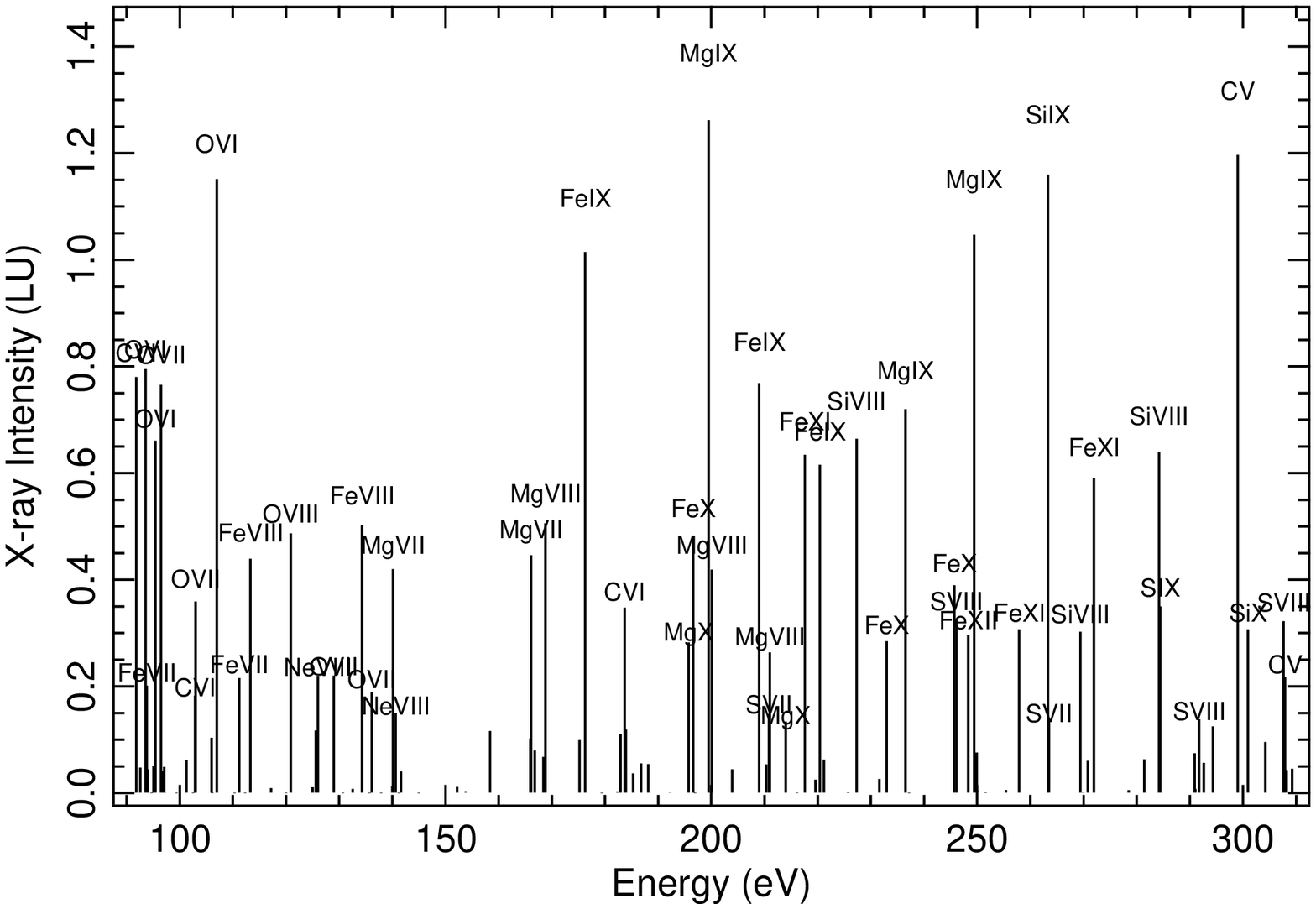}
\caption{Calculated SWCX spectra in Line Units for the 1/4 keV band. Emitting ions are marked above 
the most intense lines.}
\label{fig:spec1_4}       
\end{figure*}

For the study on the 1/4 keV band, we updated our atomic database to include Fe, Si, S, Mg ions that emit intense lines in the 0.1-0.3 keV 
range. Individual radiative transition probabilities for these ions were calculated assuming that the ions are hydrogenic. 
Moreover, photon yields were calculated using a unique neutral species, which means that no distinction 
between H and He was made. The hydrogenic ion assumption and unique neutral species choice are only approximations, 
but the results give a good estimate of the order of magnitude of the X-ray intensity for this energy range. An example 
of calculated spectra in the 0.1-0.3 keV range is presented in figure \ref{fig:spec1_4}, with the emitting ion identifying the most intense lines. 

We have calculated spectra for two LOS, upwind (UW) and downwind (DW), for an observer assumed to be at 1 AU and in a crosswind position 
with respect to the LOS. This observing geometry was chosen to match the ROSAT satellite observing constraints. In the 1/4 keV (0.1- 0.3 keV) range 
the model yields a total of 28.35 LU for the UW LOS and a total of 20.27 LU for the DW LOS. The intensity is 
probably underestimated because of the SW ion propagation in the heliosheath up to the heliopause, and in the heliotail up to $\sim$3\,000 AU, 
where all ions are used up. The outer heliospheric region is neglected (for the moment) in our model, but preliminary estimates yield a 
maximum additional $\sim$20\%\ contribution in the DW direction. In general, because of parallax effects as well, we can assume that within 
a 50\%\ error bar, these values give an good estimate of average values of the heliospheric SWCX emission in the 1/4 keV range.

\subsection{Model-Data Comparison}
\label{sec:data}
The sounding rocket survey of the University of Wisconsin and the ROSAT satellite survey have dominated the diffuse SXRB observations 
from 1972 to 1991. The Wisconsin survey allowed a full-sky mapping of the IS SXRB with a spatial resolution of $\sim$7\deg\ \citep{mccammon83}. 
The ROSAT survey improved the spatial resolution to $\sim$12' \citep{snowden97}. However, the instruments did not have any spectral resolution 
so they provided only total fluxes for each band. 

In order to compare the modeled SWCX emission level to the measured SXRB intensities at 1/4 keV, we convolve the calculated spectra 
for the UW and DW LOS with each band response in this energy range of the instruments. The bands concerning our study are B (0.13\,-\,0.188 keV), 
C (0.16\,-\,0.284 keV) for the Wisconsin survey and R1 (0.08\,-\,0.284 keV), R2 (0.14\,-\,0.284 keV) for the ROSAT survey. 
The corresponding band responses are presented in figures: 1(a) of \citet{mccammon83} for Wisconsin and 1 of \citet{snowden00} for ROSAT. 
In table \ref{tab:2} we summarize 
the total band fluxes derived from the SWCX spectra convolution with the band responses, and the measured data range for each band 
(evaluated from figures \ref{fig:BvsC} and \ref{fig:R12_Io}).


\begin{table*}
\begin{minipage}[l]{0.5\linewidth}
\begin{tabular}{lccc}
\hline\hline\noalign{\smallskip}
&\multicolumn{2}{c}{SWCX model} & Data range\\
\noalign{\smallskip}\hline\noalign{\smallskip}
Band & UW & DW & \\
\noalign{\smallskip}\hline\noalign{\smallskip}
B (cts s$^{-1}$)& 11.99 &  8.04 & [20 - 100]\\
C (cts s$^{-1}$)& 73.32 & 56.24 & [50 - 250]\\
R1 (RU $^a$)& 141.14 & 104.38 & \\
R2 (RU)& 192.07 & 147.27 &\\
R12 (RU)& 333.21 & 251.65 & [250 - 820]\\
\noalign{\smallskip}\hline
\end{tabular}
\end{minipage}
\begin{minipage}[l]{0.5\textwidth}
\caption{Summary of main band fluxes for the SWCX model in the upwind (UW) and downwind (DW) directions and measured data. For ROSAT data 
only the unabsorbed fraction I$_o$ is given. $^a$ RU = 10$^{-6}$ cts s$^{-1}$ arcmin$^{-1}$.}
\label{tab:2}       
\end{minipage}
\end{table*}

\subsubsection{Wisconsin Sounding Rocket Survey}
\label{sec:wisc}
The SXRB survey of the University of Wisconsin was performed with a series of 10 sounding rocket flights between 1972 and 1979. The 
only spectral information in the 1/4 keV range is extracted by the boron (B) and carbon (C) filters incorporated into the detector windows 
that separate the two bands B and C presented in \citet{mccammon83}, figure 1(a). 

Detailed maps derived from the complete survey were presented in \citet{mccammon83}. The maps present a clear negative correlation 
with the column density of galactic neutral hydrogen N$_{H I}$. The X-ray intensity is lowest toward the galactic plane and highest 
at high galactic latitudes. \citet{snowden90b} investigated the band intensity ratio B/C (low to high energies) and found a variation 
between 0.25 to 0.46 in a dipole-like spatial correlation, aligned with a roughly galactic center-anticenter direction, that 
they attributed to an anisotropy of the Local Hot Bubble temperature ranging between 10$^{6.2}$ and 10$^{5.9}$ K respectively.

\begin{figure}
  \includegraphics[width=0.65\textwidth]{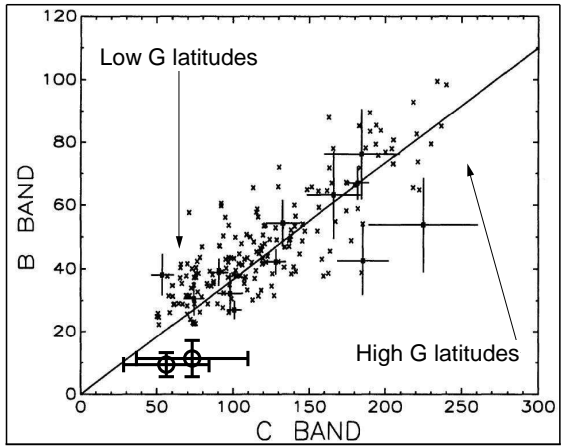}
\caption{Scatter plot of the total intensity calculated in the SWCX model (large circles) compared to total measured (crosses) 
values in B and C bands in the Wisconsin survey. The model reproduces the observed C band fluxes in the galactic plane, 
but not the B band fluxes. Figure extracted and adapted for our study from \citet{snowden90b}.}
\label{fig:BvsC}       
\end{figure}

From \citet{snowden90b} we extracted a scatter plot (fig.\ref{fig:BvsC}) comparing the B and C band measurements in the Wisconsin survey. 
The data range is summarized in table \ref{tab:2} and gives $\sim$[20-100] cts s$^{-1}$ for the B band and $\sim$[50-250] cts s$^{-1}$ for the C band. 
We superposed, in this figure, the two points (wide circles) corresponding to our simulated spectra convolved with the B and C band responses 
(see table \ref{tab:2}). Error bars are given only for information at 50\% of the SWCX band flux, in order to give an estimate of the heliospheric 
flux range in these bands.

The SWCX model in the C band predicts fluxes well within the observed values in the galactic plane (lower limits), but simulated B band flux 
is about two times lower than the lower observed limits. The band ratio B/C for the SWCX flux is accordingly inconsistent with measured values, 
and fairly constant from UW to DW directions, with a mean value of B/C $\sim$ 0.15.

\subsubsection{The ROSAT Survey}
\label{sec:rosat}
The ROSAT all-sky survey was performed in a scan mode with imaging telescopes in the soft X-ray band of 0.1\,-\,2 keV, during the first year 
of the mission. The observing geometry constraints imposed to scan the sky in a big circle, during one orbit, perpendicular 
to the spin axis that coincided with the Sun-Earth direction. Therefore a complete scan of the sky was performed in a six-month period.

The ROSAT all-sky maps in the 1/4 keV range \citep{snowden95,snowden97} confirmed the Wisconsin results and the negative correlation 
with the H\scriptsize{ I}\normalsize\ column density. The improved spatial resolution of ROSAT, with respect to previous surveys, 
allowed a better use of shadows as tools to separate the foreground from distant components of the SXRB. An extensive catalog of shadows 
in the 1/4 keV band, with fitted foreground and distant emission values was presented in \citet{snowden00}.

\begin{figure}
  \includegraphics[width=0.65\textwidth]{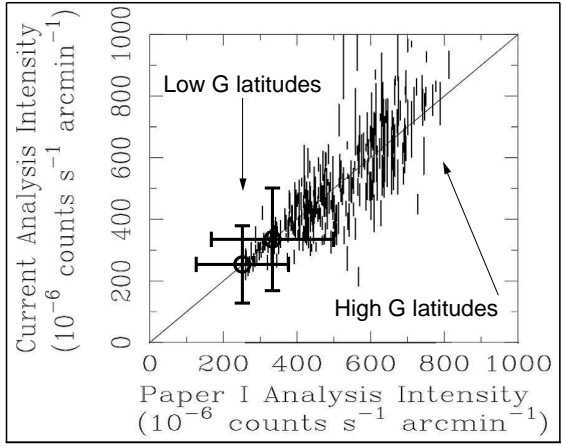}
\caption{Same as fig. \ref{fig:BvsC} but for the unabsorbed (I$_o$) emission of ROSAT shadowing observations in band R12 (R1 + R2). 
The model reproduces most of the R12 band flux in the galactic plane. Figure extracted and adapted for our study from \citet{snowden00}.}
\label{fig:R12_Io}       
\end{figure}

In figure \ref{fig:R12_Io} we present a scatter plot of foreground intensity I$_o$ measured in band R12, which is the sum of intensities 
in bands R1 and R2. Units are in 10$^{-6}$ cts s$^{-1}$ arcmin$^{-1}$, better known as ROSAT Units (RU).
The figure 
was extracted from \citet{snowden00} and the authors compare the data set analysed with two different methods. The exact comparison 
of the two methods is beyond the scope of this paper, since what we need is just the data intensity range (250-820 RU). 
In the figure y = x, thus the fitted slope is one. Simulated spectra convolved with R1 and R2 band responses and summed for the total 
result in R12 band yield 333.21 and 251.65 RU for the UW and DW direction respectively. These values, represented by the wide circles 
in fig. \ref{fig:R12_Io} with error bars at 50\%\ (to give the SWCX flux range), are within the lower limits of observational data, that correspond 
to the low galactic latitude regions.

In the same paper \citep{snowden00} confirm the dipole-like spatial correlation of the R2/R1 (high to low energies) flux ratio, 
although the variation is less pronounced than the one derived from the Wisconsin analysis, 1.25 to 1.04 from UW to DW. Our model results 
yield a constant R2/R1 ratio at an average value of 1.38.

\section{Discussion}
\label{sec:discussion}
We have presented simulation results on the SWCX soft X-ray emission in the heliosphere in the 0.1-1.0 keV energy range, that we compared 
to data from past and present observing instruments. We separated our analysis into two energy ranges: the 3/4 keV, dominated by 
the oxygen (O\scriptsize{ VII}\normalsize\ and O\scriptsize{ VIII}\normalsize ) lines and the 1/4 keV, where heavier ion lines 
as Mg, Si, S and Fe dominate the SWCX spectra. This analysis yields estimates of the SWCX heliospheric component 
within the diffuse Soft X-ray Background and confirms the large contamination of X-ray data by the heliospheric emission that needs 
to be taken into account in future determinations of the LB temperature and pressure.

The simulation results on shadowing clouds observed with Chandra, XMM-Newton and Suzaku suggest that the local 3/4 keV emission detected in front 
of shadows is entirely explained by the heliospheric SWCX emission (see table \ref{tab:1} and figure \ref{fig:newSHfit}) 
and no emission from the LB is needed at these energies. Thus, combining our results in the 3/4 keV range and previous results published 
on the 3/4 and 1/4 keV ROSAT band ratios, the strict upper limit of the LB temperature is 10$^6$ K. 
Indeed, if the LB plasma is in collisional equilibrium at 10$^6$ K, then it mainly emits in the 1/4 keV range, and only 
very little emission is produced in the oxygen lines at 3/4 keV.

But, then, how does SWCX affect the data interpretation in the 1/4 keV range? We have compared SWCX simulation results with 
data from the ROSAT and Wisconsin surveys. Both sets of measured data were originally interpreted in terms of hot gas emission in the LB 
at a temperature of $\sim$ 10$^6$ K. Variations in the observed X-ray intensity were attributed to variations in the extent 
of the emission volume and therefore the emission measure of the plasma.

SWCX emission flux calculated for the R12 and C bands (which practically cover the same energy range) yielded average values 
that explain most of the observed emission in the galactic plane, previously attributed to the restricted regions of the LB. 
Indeed, according to \citet{snowden98}, in the galactic center and anticenter directions (roughly UW and DW respectively), 
the LHB is found to have an extent of R$_{LHB}$ $\sim$ 50 pc \citep[see upper left panel of figure 10 in][]{snowden98}, which corresponds 
to an unabsorbed 1/4 keV emission of I$_o$ = R$_{LHB}$ / 0.155 $\approx$ 322 RU measured in the R12 band, while our model 
predicts $\sim$[250-333] RU. 

On the other hand, the SXRB intensity is typically brighter at higher Galactic latitudes, and SWCX cannot account for all the emission. 
This may be explained by the fact that at high latitudes the LOS's point through the chimneys allowing for non-absorbed 
halo emission to reach the observer. Moreover in the B band, which is the least overlapping in energy range with respect 
to the other 1/4 keV bands, the SWCX simulations predict about half of the observed lower limits. Besides, the SWCX model fails 
to reproduce the observed R2/R1 ratio by 10-33\%\ and the B/C ratio by 40-67\% . 

The answer, then, to the local unabsorbed SXRB puzzle might lie in the proper mix of SWCX emission 
and warm (rather than hot) LB that will fill-in the gap in the B band and reproduce the observed band ratios (Koutroumpa et al., in preparation).

\begin{acknowledgements}
DK and RL aknowledge The Institute for Theoretical Atomic, Molecular and Optical Physics (ITAMP) for travel and living expenses support 
during a visit to ITAMP facilities. The authors would like to thank Dan McCammon for providing accurate values of the effective area functions 
of the Wisconsin and ROSAT bands. We are also thankful to Steve Snowden for useful (and lively) discussions. Finally, we are really greatful 
to our referee Jeffrey Linsky for the thorough report and constructive remarks, as well as for the organisation, along 
with the organising comittee of the ISSI Local Bubble Workshop, of this most interesting meeting.
\end{acknowledgements}

\newcommand{\bibfont}{\footnotesize}
\bibliographystyle{SSRv}
\bibliography{bibDSSRv}   


\end{document}